\def\EMPTYSET{\mbox{\O}} 

\documentstyle[twoside,psfig]{article}

\catcode`\@=11
\long\def\@makefntext#1{
\protect\noindent \hbox to 3.2pt {\hskip-.9pt  
$^{{\eightrm\@thefnmark}}$\hfil}#1\hfill}		

\def\thefootnote{\fnsymbol{footnote}}
\def\@makefnmark{\hbox to 0pt{$^{\@thefnmark}$\hss}}	
	
\def\ps@myheadings{\let\@mkboth\@gobbletwo
\def\@oddhead{\hbox{}
\rightmark\hfil\eightrm\thepage}   
\def\@oddfoot{}\def\@evenhead{\eightrm\thepage\hfil
\leftmark\hbox{}}\def\@evenfoot{}
\def\sectionmark##1{}\def\subsectionmark##1{}}

\oddsidemargin=\evensidemargin
\renewcommand{\thefootnote}{\fnsymbol{footnote}}

\newcounter{sectionc}\newcounter{subsectionc}\newcounter{subsubsectionc}
\renewcommand{\section}[1] {\vspace{12pt}\addtocounter{sectionc}{1} 
\setcounter{subsectionc}{0}\setcounter{subsubsectionc}{0}\noindent 
	{\tenbf\thesectionc. #1}\par\vspace{5pt}}
\renewcommand{\subsection}[1] {\vspace{12pt}\addtocounter{subsectionc}{1} 
	\setcounter{subsubsectionc}{0}\noindent 
	{\bf\thesectionc.\thesubsectionc. {\kern1pt \bfit #1}}\par\vspace{5pt}}
\renewcommand{\subsubsection}[1] {\vspace{12pt}\addtocounter{subsubsectionc}{1}
	\noindent{\tenrm\thesectionc.\thesubsectionc.\thesubsubsectionc.
	{\kern1pt \tenit #1}}\par\vspace{5pt}}
\newcommand{\nonumsection}[1] {\vspace{12pt}\noindent{\tenbf #1}
	\par\vspace{5pt}}

\newcounter{appendixc}
\newcounter{subappendixc}[appendixc]
\newcounter{subsubappendixc}[subappendixc]
\renewcommand{\thesubappendixc}{\Alph{appendixc}.\arabic{subappendixc}}
\renewcommand{\thesubsubappendixc}
	{\Alph{appendixc}.\arabic{subappendixc}.\arabic{subsubappendixc}}

\renewcommand{\appendix}[1] {\vspace{12pt}
        \refstepcounter{appendixc}
        \setcounter{figure}{0}
        \setcounter{table}{0}
        \setcounter{lemma}{0}
        \setcounter{theorem}{0}
        \setcounter{corollary}{0}
        \setcounter{definition}{0}
        \setcounter{equation}{0}
        \renewcommand{\thefigure}{\Alph{appendixc}.\arabic{figure}}
        \renewcommand{\thetable}{\Alph{appendixc}.\arabic{table}}
        \renewcommand{\theappendixc}{\Alph{appendixc}}
        \renewcommand{\thelemma}{\Alph{appendixc}.\arabic{lemma}}
        \renewcommand{\thetheorem}{\Alph{appendixc}.\arabic{theorem}}
        \renewcommand{\thedefinition}{\Alph{appendixc}.\arabic{definition}}
        \renewcommand{\thecorollary}{\Alph{appendixc}.\arabic{corollary}}
        \renewcommand{\theequation}{\Alph{appendixc}.\arabic{equation}}
        \noindent{\tenbf Appendix \theappendixc #1}\par\vspace{5pt}}
\newcommand{\subappendix}[1] {\vspace{12pt}
        \refstepcounter{subappendixc}
        \noindent{\bf Appendix \thesubappendixc. {\kern1pt \bfit #1}}
	\par\vspace{5pt}}
\newcommand{\subsubappendix}[1] {\vspace{12pt}
        \refstepcounter{subsubappendixc}
        \noindent{\rm Appendix \thesubsubappendixc. {\kern1pt \tenit #1}}
	\par\vspace{5pt}}

\newcommand{\textlineskip}{\baselineskip=13pt}
\newcommand{\smalllineskip}{\baselineskip=10pt}

\def\eightcirc{
\begin{picture}(0,0)
\put(4.4,1.8){\circle{6.5}}
\end{picture}}
\def\eightcopyright{\eightcirc\kern2.7pt\hbox{\eightrm c}} 

\newcommand{\copyrightheading}[1]
	{\vspace*{-2.5cm}\smalllineskip{\flushleft
	{\footnotesize Modern Physics Letters A, #1}\\
	{\footnotesize $\eightcopyright$\, World Scientific Publishing
	 Company}\\
	 }}

\newcommand{\publisher}[2]{{\begin{center}\footnotesize\smalllineskip 
	Received #1\\
	Revised #2
	\end{center}
	}}

\def\abstracts#1#2#3{{
	\centering{\begin{minipage}{4.5in}\footnotesize\baselineskip=10pt
	\parindent=0pt #1\par 
	\parindent=15pt #2\par
	\parindent=15pt #3
	\end{minipage}}\par}} 

\newcommand{\bibbf}{\ninebf}
\renewenvironment{thebibliography}[1]
	{\frenchspacing
	 \ninerm\baselineskip=11pt
	 \begin{list}{\arabic{enumi}.}
        {\usecounter{enumi}\setlength{\parsep}{0pt}     
	 \setlength{\leftmargin 12.7pt}{\rightmargin 0pt} 
         \setlength{\itemsep}{0pt} \settowidth
	{\labelwidth}{#1.}\sloppy}}{\end{list}}

\newcounter{itemlistc}
\newcounter{romanlistc}
\newcounter{alphlistc}
\newcounter{arabiclistc}

\newcommand{\fcaption}[1]{
        \refstepcounter{figure}
        \setbox\@tempboxa = \hbox{\footnotesize Fig.~\thefigure. #1}
        \ifdim \wd\@tempboxa > 5in
           {\begin{center}
        \parbox{5in}{\footnotesize\smalllineskip Fig.~\thefigure. #1}
            \end{center}}
        \else
             {\begin{center}
             {\footnotesize Fig.~\thefigure. #1}
              \end{center}}
        \fi}

\newcommand{\tcaption}[1]{
        \refstepcounter{table}
        \setbox\@tempboxa = \hbox{\footnotesize Table~\thetable. #1}
        \ifdim \wd\@tempboxa > 5in
           {\begin{center}
        \parbox{5in}{\footnotesize\smalllineskip Table~\thetable. #1}
            \end{center}}
        \else
             {\begin{center}
             {\footnotesize Table~\thetable. #1}
              \end{center}}
        \fi}

\def\@citex[#1]#2{\if@filesw\immediate\write\@auxout
	{\string\citation{#2}}\fi
\def\@citea{}\@cite{\@for\@citeb:=#2\do
	{\@citea\def\@citea{,}\@ifundefined
	{b@\@citeb}{{\bf ?}\@warning
	{Citation `\@citeb' on page \thepage \space undefined}}
	{\csname b@\@citeb\endcsname}}}{#1}}

\newif\if@cghi
\def\cite{\@cghitrue\@ifnextchar [{\@tempswatrue
	\@citex}{\@tempswafalse\@citex[]}}
\def\citelow{\@cghifalse\@ifnextchar [{\@tempswatrue
	\@citex}{\@tempswafalse\@citex[]}}
\def\@cite#1#2{{$\null^{#1}$\if@tempswa\typeout
	{IJCGA warning: optional citation argument 
	ignored: `#2'} \fi}}

\def\pmb#1{\setbox0=\hbox{#1}
	\kern-.025em\copy0\kern-\wd0
	\kern.05em\copy0\kern-\wd0
	\kern-.025em\raise.0433em\box0}

\def\fnt#1#2{\footnotetext{\kern-.3em
	{$^{\mbox{\scriptsize #1}}$}{#2}}}

\def\fpage#1{\begingroup
\voffset=.3in
\thispagestyle{empty}\begin{table}[b]\centerline{\footnotesize #1}
	\end{table}\endgroup}

\def\runninghead#1#2{\pagestyle{myheadings}
\markboth{{\protect\footnotesize\it{\quad #1}}\hfill}
{\hfill{\protect\footnotesize\it{#2\quad}}}}
\font\tenrm=cmr10
\font\tenit=cmti10 
\font\tenbf=cmbx10
\font\bfit=cmbxti10 at 10pt
\font\ninerm=cmr9

\font\ninebf=cmbx9
\font\eightrm=cmr8

\def\qed{\hbox{${\vcenter{\vbox{			
   \hrule height 0.4pt\hbox{\vrule width 0.4pt height 6pt
   \kern5pt\vrule width 0.4pt}\hrule height 0.4pt}}}$}}

\renewcommand{\thefootnote}{\fnsymbol{footnote}}	

\begin{document}
\setlength{\textheight}{7.7truein}  

\runninghead{R. B. Salgado}
{Some Identities for the Quantum Measure and its Generalizations}

\normalsize\textlineskip
\thispagestyle{empty}
\setcounter{page}{1}

\copyrightheading{}			

\vspace*{0.88truein}

\fpage{1}
\centerline{\bf SOME IDENTITIES FOR THE QUANTUM MEASURE}
\baselineskip=13pt
\centerline{\bf  AND ITS GENERALIZATIONS}
\vspace*{0.37truein}
\centerline{\footnotesize ROBERTO B. SALGADO\footnote{E-mail: salgado@physics.syr.edu} }
\baselineskip=12pt
\centerline{\footnotesize\it Department of Physics, Syracuse University}
\baselineskip=10pt
\centerline{\footnotesize\it Syracuse, New York, 13244,
USA}
\vspace*{10pt}

\publisher{(received date)}{(revised date)}

\vspace*{0.21truein}
\abstracts{After a brief review of classical probability theory (measure theory), 
we present an observation (due to Sorkin) concerning 
an aspect of probability in quantum mechanics.
Following Sorkin, we introduce a generalized measure theory
based on a hierarchy of ``sum-rules.'' The first sum-rule 
yields classical probability theory, and the second yields
a generalized probability theory that includes 
quantum mechanics as a special case.
We present some algebraic relations involving these sum-rules.
This may be useful for the study of the higher-order sum-rules
and possible generalizations of quantum mechanics.
We conclude with some open questions and suggestions for further work.
}{}{}

\setcounter{footnote}{0}
\renewcommand{\thefootnote}{\alph{footnote}}



\vspace*{1pt}\textlineskip	
\section{Introduction}	
\vspace*{-0.5pt}
\noindent
One could take the point of view that what is at the essence of quantum
mechanics is the failure of ``the classical additivity of probabilities,''
as demonstrated by the famous two-slit experiment.

Consider the set $C$ of all electron worldlines (histories)
that leave the emitter at a given instant and arrive at a 
particular detector on the other side of a double-slit screen 
at a later instant.  
Suppose we block off only the second slit.
Let $A$ be the subset of those worldlines which pass through the first slit.
Similarly, we block off only the first slit and let $B$ be the subset of 
those worldlines which pass through the second slit.  
Ignoring the possibility of the electron 
winding around so that it passes through both slits, 
we have $C=A\sqcup B$, where $\sqcup$ denotes disjoint-union.
This suggests 4 ($=2^2$) experimental combinations of $2$ disjoint
alternatives: ``both slits open,''
``only slit-A open,'' ``only slit-B open,'' and ``no slits open.''

Classical measure theory (probability theory) assigns to each 
measurable set $X$ of histories a non-negative number $P(X)$. 
So, we can ask about the validity of the ``sum rule''
$$P(A \sqcup B) \stackrel{?}{=} P(A)+P(B),$$
which we write as
$$ P(A) + P(B) - P(A \sqcup B)\stackrel{?}{=} 0.$$
Implicitly, we assume that $P(\EMPTYSET)=0$.

Of course, 
the probability function $P$ in classical physics assigns to a set $A$
a positive additive quantity, $a_1$,
which we call the ``classical amplitude.''
So,
$$P(A)= a_1$$
$$P(B)= b_1$$
$$P(A \sqcup B) = a_1 +b_1. $$
We verify that the sum-rule is satisfied
$$ P(A) + P(B) - P(A \sqcup B)=
(a_1) + (b_1)-(a_1+b_1) =0.$$

However, 
the probability function $P_2$ in quantum physics assigns to a set $A$
a sum of the square-norms of additive quantities, $a_2$. 
(Refer to \cite{QMT,QMTI} for details.)
We call these additive quantities the ``quantum amplitudes.''
So,
$$P_2(A)= \sum_i a_{i,2}{}^* a_{i,2}$$
$$P_2(B)= \sum_i b_{i,2}{}^* b_{i,2}$$
$$P_2(A \sqcup B) = \sum_i(a_{i,2} +b_{i,2})^*(a_{i,2}+b_{i,2}) $$
The corresponding sum-rule, however, fails
\begin{eqnarray*}
 \lefteqn{P_2(A) + P_2(B) - P_2(A \sqcup B)}\\
&=& \sum_i a_{i,2}{}^* a_{i,2} + \sum_i b_{i,2}{}^* b_{i,2}
-\sum_i(a_{i,2} +b_{i,2})^* (a_{i,2}+b_{i,2})\\
&=& -\sum_{i} \left( a_{i,2}{}^*b_{i,2}+b_{i,2}{}^*a_{i,2}  \right)\\
&\neq& 0.
\end{eqnarray*}
This is the failure of the 
additivity of probabilities in quantum mechanics.

Let us define the ``interference term''\footnote{%
This expression for $I_2$ is {\em minus} the definition given in \cite{QMT}.
In general, our expressions for $I_{even}$ differ in sign.
Our convention is chosen for mathematical convenience.
}
$$I_2(A,B) \equiv  P(A) + P(B)-P(A \sqcup B),$$
which measures the failure of the additivity of probabilities.
Then, we can say that classical probabilities have the property that ``$I_2=0$'',
and that quantum probabilities have the property that ``$I_2\neq 0$.''

Following Sorkin,\cite{QMT} consider next the not-so-famous 
three-slit experiment, with
8 ($=2^3$) experimental combinations of $3$ disjoint alternatives.
Consider the following function
\begin{eqnarray*}
I_3(A,B,C)
&\equiv& P(A) + P(B) + P(C) 
   -P(A \sqcup B) - P(A \sqcup C) - P(B \sqcup C)\\ 
& &+ P(A \sqcup B \sqcup C).
\end{eqnarray*}
With classical probabilities, this evaluates to zero.
\begin{eqnarray*}
I_3(A,B,C)
&=&(a_1) +(b_1)+ (c_1) - 
(a_1+b_1) - (a_1+c_1) - (b_1+c_1)\\ 
& &+(a_1+b_1+c_1)\\
&=&0
\end{eqnarray*}
This result can be expected from a simple application of classical measure
theory, as we show in the next section.

With quantum probabilities, this, surprisingly, also evaluates to zero! 
\begin{eqnarray*}
\lefteqn{I_3(A,B,C) }\\
&=&\sum_i a_{i,2}{}^* a_{i,2} 
   + \sum_i b_{i,2}{}^* b_{i,2} 
   + \sum_i c_{i,2}{}^* c_{i,2} \\
& & -\sum_i (a_{i,2}+b_{i,2})^* (a_{i,2}+b_{i,2}) 
	-\sum_i (a_{i,2}+c_{i,2})^* (a_{i,2}+c_{i,2})\\
& &\qquad\qquad
	-\sum_i (b_{i,2}+c_{i,2})^* (b_{i,2}+c_{i,2})\\ 
& &+\sum_i (a_{i,2}+b_{i,2}+c_{i,2})^* (a_{i,2}+b_{i,2}+c_{i,2})\\
&=&0
\end{eqnarray*}

We can say that, for disjoint alternatives, 
classical probabilities have the property that ``$I_2=0$ and $I_3=0$,''
and quantum probabilities have the property that ``$I_2 \neq 0$ and $I_3=0$.''

This was Sorkin's observation concerning an aspect of probability
in quantum mechanics.  It seems to say that quantum probabilities reveal
themselves as a rather mild generalization of classical probabilities in the
sense that the probability sum-rules are only slightly different:
additivity is lost for two disjoint alternatives but not three.
Such a generalization allows one to work directly with the 
quantum probabilities (quantum measures) instead of 
indirectly with the quantum amplitudes 
(wavefunctions defined over a spacelike hypersurface).
This generalized viewpoint is probably needed to formulate,
say, a quantum theory of causal sets,\cite{SAACS,Forks}
or any other quantum theory of gravity
which does not naturally have hypersurfaces.

\section{Classical Measure Theory}
\noindent
In this section, we review classical measure theory (\'a la Kolmogorov).
A measure space $({\cal S},M,|\cdot|)$
consists of a set ${\cal S}$, a collection $M$ of certain subsets of ${\cal S}$ (called
``measurable sets''),
and a function (called the ``measure'') 
$|\cdot|:M\rightarrow R^+$, where $R^+=[0,\infty)$,
such that
\begin{enumerate}
\item $\EMPTYSET\in M$, and $|\EMPTYSET|=0$\\
	``the empty set is measurable and has measure zero''
\item ${\cal S}\in M$, and $|{\cal S}|=1$\\
	``the universal set is measurable and has measure one''
\item if $A\in M$, then $A^c \in M$\\
	``the complement of a measurable set is also measurable''
\item for $A_1, A_2, \ldots \in M$, we have $\cup A_i \in M$\\
	``the union of a countable collection of measurable sets 
	is also measurable''
\item for mutually-disjoint $S_1, S_2, \ldots \in M$, we have
	$ |\sqcup  S_i| =  \sum | S_i|$\\
	``the disjoint-union of a countable collection of mutually-disjoint
	measurable sets is also measurable,
	and its measure is the sum of the individual measures''\\
	\label{enum:measure-disjoint}
\end{enumerate}

Axiom~\ref{enum:measure-disjoint} permits a ``frequency'' or ``area''
interpretation for classical measure theory.
In particular, this condition directly yields
\begin{equation}
|A \sqcup B| = |A| + |B|,
\end{equation}
(which is  ``$I_2=0$'') and
\begin{equation}
|A \sqcup B \sqcup C| = |A| + |B| + |C|.
\end{equation}
By reapplying this axiom and using the associativity
and commutativity of $\sqcup$, we obtain 
for the case of
three mutually disjoint sets:
$$\hspace*{-0.25in}
\begin{array}{rcrrrrrr}
|A \sqcup (B \sqcup C)| &=& |A|&    &    &            &+|B\sqcup C|&            \\
|(A \sqcup B)\sqcup C| &=&    &    & |C|&+|A\sqcup B|&            &             \\
|B \sqcup (C \sqcup A)| &=&    & |B|&    &            &            &+|C\sqcup A| \\
-2|A \sqcup B \sqcup C| &=&-2|A|&-2|B|&-2|C|&
\vspace*{1ex}\\
\hline  \vspace*{1ex} 
|A \sqcup B \sqcup C| &=&-|A|&-|B|&-|C|&+|A \sqcup B|&+|B \sqcup C|&+|C \sqcup A|
\end{array}
$$
(which is  ``$I_3=0$'').
\vspace*{4ex}

For later comparison, let us define the ``generalized interference term''
for any collection of mutually-disjoint subsets $S_1, S_2, \cdots \in M$:
\begin{eqnarray}
\lefteqn{I_n(S_1,S_2,S_3,\cdots,S_n)}\nonumber\\
&\equiv&
\sum_i |S_i|\
- \sum_{\mbox{distinct }i,j} |S_i \sqcup S_j|
+  \sum_{\mbox{distinct }i,j,k} |S_i \sqcup S_j \sqcup S_k|\ \mp \cdots  \nonumber\\
&&\qquad
-(-1)^{n-1} \sum_i |S_1 \sqcup S_2 \sqcup \cdots \sqcup \stackrel{omit}{(S_i)}\sqcup \cdots \sqcup S_n|
\nonumber\\
&&\qquad
-(-1)^n |S_1 \sqcup S_2 \sqcup \cdots \sqcup S_n|,\label{INTERFERENCE}
\end{eqnarray}
where each $I_n$ is a real-valued symmetric set-function on 
$n$ mutually-disjoint measurable sets.  Clearly, the vanishing of the $n^{th}$
generalized interference term encodes the $n^{th}$-order sum-rule.

Then,
axiom~\ref{enum:measure-disjoint} (for finite sums) may be re-expressed by
\begin{enumerate}
\item[\ref{enum:measure-disjoint}.] 
For all $n \geq 2$, and 
for any collection of mutually-disjoint subsets $S_1, S_2, \cdots, S_n\in M$,
we have $I_n(S_1,S_2,\cdots,S_n)=0$, 
which yields
\begin{eqnarray*}
\lefteqn{|S_1 \sqcup S_2 \sqcup \cdots \sqcup S_n|}\\
&\equiv&
(-1)^n \left(
\sum_i |S_i|\
- \sum_{\mbox{distinct }i,j} |S_i \sqcup S_j|\right.\\
&&\qquad\qquad
+  \sum_{\mbox{distinct }i,j,k} |S_i \sqcup S_j \sqcup S_k|\ \mp \cdots  \\
&&\qquad\qquad\left.
-(-1)^{n-1}\sum_i |S_1 \sqcup S_2 \sqcup \cdots \sqcup \stackrel{omit}{(S_i)}\sqcup \cdots \sqcup S_n|
\right).
\end{eqnarray*}
\end{enumerate}
This will be generalized in the next section.

Let us conclude this section with the following interesting fact.
Let $A$ and $B$ be disjoint subsets, and let $a$ and $b$ denote their amplitudes, respectively. 
Since amplitudes are additive, the amplitude of, say, their disjoint union $A\sqcup B$ is $a+b$.
Let $r$ be a non-negative integer and let $P_r$ denote the probability function 
which assigns to any disjoint union of subsets
$A\sqcup B \sqcup \cdots \sqcup M$ the $r^{th}$-power of its amplitude
$(a+b +\ldots+m)^r$.
Then, it will be shown that if $r<n$, then $I_n=0$. 

In order to see this, let us express $I_n$ in terms of $P_r$:
\begin{eqnarray}
\lefteqn{I_n(S_1,S_2,S_3,\cdots,S_n)}\nonumber\\
&\equiv&
\sum_i (s_i)^r\
- \sum_{\mbox{distinct }i,j} (s_i + s_j)^r
+  \sum_{\mbox{distinct }i,j,k} (s_i + s_j +s_k)^r\ \mp \cdots  \nonumber\\
&&\qquad
-(-1)^{n-1} \sum_i (s_1 + s_2 + \ldots + \stackrel{omit}{(s_i)} + \ldots + s_n)^r
\nonumber\\
&&\qquad
-(-1)^n (s_1 + s_2 + \ldots + s_n)^r.\label{INTERFERENCE2}
\end{eqnarray}
Note that every term has degree $r$.
The strategy is to consider any term of the form 
$(s_i)^{r_i} (s_j)^{r_j}\cdots (s_\ell)^{r_\ell}$, 
where $r_i,\ r_j,\ \ldots,\ r_\ell$ are positive integers whose sum is $r$,
and show that its coefficient in $I_n$ vanishes for $r<n$.

Consider such a term $(s_i)^{r_i} (s_j)^{r_j}\cdots (s_\ell)^{r_\ell}$, 
which involves $\ell$ of $n$ possible (atomic) amplitudes.
Observe that this term occurs with coefficient $r!/(r_i ! r_j ! \cdots r_\ell !)$
in the $r^{th}$-power of any multinomial which contains 
$(s_i+ s_j+ \ldots+ s_\ell)$.
From all of the multinomials with exactly $m$ terms, raised to the $r^{th}$ power, 
i.e., $(s_i+ s_j + \ldots + s_\ell + \ldots s_m)^r$,
the term $(s_i)^{r_i} (s_j)^{r_j}\cdots (s_\ell)^{r_\ell}$ appears 
$\scriptsize\left( \begin{array}{cc} n-\ell \\ m-\ell \end{array}\right)=(n-\ell)!/( (m-\ell)! (n-m)! )$
times. So, the sum of the coefficients of the term $(s_i)^{r_i} (s_j)^{r_j}\cdots (s_\ell)^{r_\ell}$
in $I_n$ is
$$\left(-\sum_{m=1}^n (-1)^m \left( \begin{array}{cc} n-\ell \\ m-\ell \end{array}\right)\right)
\frac{r!}{r_i ! r_j ! \cdots r_\ell !}.
$$
Note that
\begin{eqnarray*}
\sum_{m=1}^n (-1)^m \left( \begin{array}{cc} n-\ell \\ m-\ell \end{array}\right)
&=&
\sum_{m=\ell}^n (-1)^m \left( \begin{array}{cc} n-\ell \\ m-\ell \end{array}\right)\\
&=&
\sum_{p=0}^{n-\ell} (-1)^{p+\ell} \left( \begin{array}{cc} n-\ell \\ p \end{array}\right).
\end{eqnarray*}

Since $0\leq r_i + r_j + \ldots + r_\ell = r$, with each of 
$r_i,\ r_j, \ldots,\ r_\ell >0$, we have that $\ell$ ranges from $0$ to $r$.\footnote{%
	If $\ell=0$, then the sum of the $r_i$ is empty and, so, $r=0$.
} \
Now, suppose $r<n$.  Thus, we have $\ell < n$, and the sum and, therefore, $I_n$ is zero.\footnote{%
	Using $(1+x)^k=\sum_{i=0}^k (-1)^i \!\left(\hspace*{-1.5ex} \begin{array}{c} k\\i \end{array} \hspace*{-1.5ex}\right)x^k$
	with $k>0$ and $x=-1$, we find that the sum is equal to $0$.
}

For the cases $r \geq n$, the only surviving terms are those for which $\ell=n$. So,
for $r=n$, we have $I_n=-(-1)^n n! s_1 s_2 \cdots s_n$.
For $r>n$, we have 
$I_n=-(-1)^n \sum \frac{r!}{r_1! r_2! \cdots r_n !} (s_1)^{r_1} (s_2)^{r_2}\cdots (s_n)^{r_n}$,
where the sum is over all positive-integer partitions of $r$ into $n$ parts.

\section{Generalized Measure Theory}
\noindent
Following Sorkin, we make a replacement of 
axiom~\ref{enum:measure-disjoint}.

\begin{enumerate}
\item[\ref{enum:measure-disjoint}$'$.] 
There exists an $n\geq 2,$
such that, for any collection of mutually-disjoint subsets $S_1, S_2, \cdots , S_n\in M$,
we have
$I_n(S_1,S_2,\cdots,S_n)= 0$
but
$I_{n-1}(S_1,S_2, \cdots, \stackrel{omit}{(S_j)}\nolinebreak,\linebreak \cdots ,S_n)\neq 0
$, in general.\\
\end{enumerate}
\noindent
This encodes the requirement that the Generalized Measure satisfies the
$n^{th}$-order sum-rule but not the $(n-1)^{st}$ sum-rule.\\

A concomitant of this new axiom is the following lemma:\\

\noindent{\bf Lemma~1. }\footnote{%
This expression is {\em minus} the corresponding expression given in \cite{QMT}.
This provides consistency with the sign-convention we chose earlier.
} 
\begin{eqnarray*}
\lefteqn{I_n(A,B,C,\cdots,N)=}&&\\
&&
I_{n-1}(A, C, \cdots, N) + I_{n-1}(B, C, \cdots, N)
-I_{n-1}(A\sqcup B, C, \cdots, N).
\end{eqnarray*}

It is easy to verify this for $n=2$ and $n=3$.

For $n=2$, which is at the level of classical measure theory,
\begin{eqnarray*}
I_2(A,B)&\stackrel{?}{=}&I_1(A) + I_1(B) - I_1(A\sqcup B)  \\
	&\stackrel{\surd}{=}& |A| + |B| - |A\sqcup B| ,
\end{eqnarray*}
where we used the definition of $I_1$ in the last step.

For $n=3$, which is at the level of ``Quantum Measure theory,''
\begin{eqnarray*}
I_3(A,B,C)&\stackrel{?}{=}& I_2(A,C) + I_2(B,C) - I_2(A\sqcup B,C) \\
	&\stackrel{?}{=}&\left( |A| + |C| - |A\sqcup C| \right)
	    +\left( |B| + |C| - |B\sqcup C| \right)\\
	& & \qquad -\left( |A\sqcup B| + |C| - |(A\sqcup B)\sqcup C| \right)\\
	&\stackrel{\surd}{=}& |A| + |B| + |C| - |A\sqcup B|
	    - |B\sqcup C| 
	    - |C\sqcup A|\\
	&&\qquad + |A\sqcup B\sqcup C| ,
\end{eqnarray*}
where we used the definition of $I_2$ in the second step.

We will give a full proof for all $n$ later in Section~5.  
For now, let us see what the lemma implies.

First, for any collection of disjoint subsets,
$$\mbox{if } I_{n-1}=0, \quad\mbox{ then } I_n=0,$$
and therefore $I_{n+1}=0$, $I_{n+2}=0$, etc.
That is, the vanishing of $I_n$ for some $n$ 
implies that all higher functions
vanish.  So, one can characterize classical probability as an 
``$I_2=0$'' Generalized Measure theory, and quantum probability
as an ``$I_3=0$'' Generalized Measure theory.

Secondly, 
$$\mbox{if } I_n=0, \quad\mbox{ then } I_{n-1} \mbox{ is additive in its arguments}.$$
For the ``$I_3=0$'' theories, this means that $I_2$ is a bi-additive
function.
This suggests the quadratic relation between amplitudes and probabilities.
In fact, Sorkin\cite{QMT} used this relation to show that every Quantum Measure
comes from an extension of the 
$I_2$ function applied to general (non-disjoint) arguments.

Another concomitant of 
axiom~\ref{enum:measure-disjoint}$'$ \ is the following lemma:\\

\noindent{\bf Lemma~2.} 
\begin{eqnarray*}
\lefteqn{
\left|A\sqcup B \sqcup C \sqcup \cdots \sqcup N \right|=}\\
&&\sum_i I_1(S_i)
-
\sum_{\mbox{distinct }i,j} I_2(S_i,S_j)
+
\sum_{\mbox{distinct }i,j,k} I_3(S_i,S_j,S_k)
\mp \ \cdots\ \\
&&
-(-1)^{n-1}\sum_i I_{n-1}(A, B,\cdots, \stackrel{omit}{S_i}, \cdots, N)\\
& &- (-1)^n  I_n(A, B, C, \cdots, N).
\end{eqnarray*}
This expresses the Generalized Measure of a disjoint union of a finite collection
of subsets in terms of the generalized interferences among all of these subsets.
This expression resembles the definition of the generalized interference term
(see Eq.~(\ref{INTERFERENCE})).  Indeed, we will show that there is a kind of duality
relation between the two, and this will be used in Section~5 
to give a complete proof of this lemma.

\section{An Algebraic Formulation}\label{alg}
\subsection{The Ring $Z{\cal P(S)}$}
\noindent
Consider a set ${\cal S}$ (of histories) and the set ${\cal P(S)}$ 
of all of its subsets [i.e., the power set of $\cal S$].
In practice, one would only use a smaller collection $M$ of 
measurable subsets which is at least closed under disjoint union
and contains the empty set. 

We wish to define the set $Z{\cal P(S)}$ of finite ``formal linear combinations'' 
of the elements of ${\cal P(S)}$ with integer coefficients 
[i.e., the free module on ${\cal P(S)}$ over the integers $Z$].
A typical element of $Z{\cal P(S)}$ is of the form $\sum n_i S_i$, where
$S_i \in {\cal P(S)}$ and $n_i \in Z$, of which only a finite number
are nonzero.  
We denote the additive-identity (``zero'')  by ``$0$''.
For clarity, we write ``$A$'' for the element ``$1A$'' and ``$-A$''
for its additive inverse ``$(-1)A$''.

We endow this set with a multiplication rule 
$\cdot: Z{\cal P(S)} \times Z{\cal P(S)} \rightarrow Z{\cal P(S)}$
by
\begin{eqnarray}
\left( \sum_i a_i S_i\right)\cdot \left( \sum_j b_j S_j\right)
&:=&\left( \sum_k p_k S_k\right)
\label{mult}
\end{eqnarray}
where 
\begin{eqnarray*}
p_k=
\displaystyle \sum_{  \stackrel{i,j} { S_i\sqcup S_j = S_k } }
					  \hspace{-2ex}a_i b_j
\end{eqnarray*}
with the understanding that we take $p_k=0$ if the summation is empty.
This rather complicated definition is a generalization of the product of 
two monomials
\begin{eqnarray*}
A \cdot B = 
(A\sqcup B).
\end{eqnarray*}

By definition, the empty set $\EMPTYSET \in \cal P(S)$ 
is disjoint with every element of $\cal P(S)$. So, it makes
sense to form  $A \sqcup \EMPTYSET$, which evaluates to $A$.
Thus, it follows that the multiplicative-identity (``unit'') is 
``$\EMPTYSET$'':
\begin{eqnarray*}
A\cdot \EMPTYSET &=& \EMPTYSET \cdot A\ =\ A.
\end{eqnarray*}
This should not be confused with the fact that ``products with zero are zero'':
\begin{eqnarray*}
A\cdot 0 &=& 0\cdot A \ =\ 0.
\end{eqnarray*}

This multiplication rule is obviously commutative. 
In the appendix, we prove the associativity and distributivity over addition,
which shows that $(Z{\cal P(S)},+,\cdot)$ is a ring.

\pagebreak

\subsection{The Circle-Product}
\noindent 
We now define a multiplication rule on $Z{\cal P(S)}$, 
called the ``circle-product'' or ``circle composition operator.''\footnote{
Such an operator is used to define the Jacobson radical of a ring. See reference~\cite{JAC}.
}~  
For any pair of subsets $A,B \in {\cal P(S)}$,
$$ (1A) \circ (1B) := (1A) +(1B) +(-1)(A\cdot B) $$
or, simply,
$$ A \circ B := A +B -A\cdot B.$$
Note that
\begin{eqnarray*} 
0 \circ B&=& 0 + B -0\cdot B\\ &=& 0 + B - 0\\ &=& B
\end{eqnarray*}\vspace*{-3ex}
but
\begin{eqnarray*}
\EMPTYSET \circ B &=& \EMPTYSET + B - \EMPTYSET \cdot B\\ 
&=& \EMPTYSET + B -B\\ &=&  \EMPTYSET.
\end{eqnarray*}
So, viewing the circle-product multiplicatively, 
the ``circle-identity''
coincides with the additive-identity $0$, and
the ``circle-zero'' coincides with the multiplicative-identity $\EMPTYSET$.\footnote{
	Note that if we had defined the circle-operator with the opposite sign-convention,
$A \circ B = A\cdot B - A - B$,
	then we would have had 
$0 \circ B = 0\cdot B - 0 - B = -B$ and 
$\EMPTYSET \circ B = \EMPTYSET\cdot B - \EMPTYSET - B = -\EMPTYSET$.
}%

Clearly, the circle-product is commutative.
Associativity of the circle-product, however, arises in a nontrivial way from 
that of the multiplication rule.

First, observe that the circle-product generally does not distribute over addition.
In fact, this is a consequence of the distributivity of multiplication over addition.
\begin{eqnarray*}
(A + B) \cdot C  &{=}& A\cdot C + B \cdot C\\
(A + B) + C  - (A+B)\circ C &=& A+C-A\circ C + B+C -B\circ C\\
             - (A+B)\circ C &=& -A\circ C -B\circ C  +C\\
               (A+B)\circ C &\neq& A\circ C +B\circ C.
\end{eqnarray*}
Instead, the circle-product is said to be 
``quasi-distributive''\cite{HLC,MEN} over addition:
$$(A + B) \circ C  = A\circ C + B \circ C - C.$$
This implies, for example, that
\begin{eqnarray*} 
(nA)\circ C 
&=& n(A\circ C) - (n-1)C\\
&\neq& n(A\circ C).
\end{eqnarray*}

However, 
the circle-product does distribute over ``affine sums,'' 
i.e., sums whose coefficients add up to $1$.
\begin{eqnarray*}
\left(\sum n_i S_i\right) \circ X &\stackrel{?}{=}& \sum n_i\left(  S_i\circ X \right)\\
\left(\sum n_i S_i\right) +  X - \left(\sum n_i S_i\right)\cdot X 
&\stackrel{?}{=}& \sum  n_i ( S_i + X - S_i\cdot  X) \\
\sum (n_i S_i) +  X - \sum (n_i S_i\cdot X) &\stackrel{?}{=}& \sum ( n_i S_i ) + \left(\sum  n_i\right) X  - \sum ( n_i S_i\cdot X )\\
X &\stackrel{?}{=}&  \left(\sum  n_i\right)X\\
1 &\stackrel{\surd}{=}&  \left(\sum  n_i\right).
\end{eqnarray*}
Such a condition could be characterized as ``affine distributivity.''\cite{RDS}

In particular, this implies that 
\begin{eqnarray*}
(A + B - A\cdot B) \circ C  &=& A\circ C + B\circ C - (A\cdot B)\circ C
\end{eqnarray*}
or, using the definition of the circle-product,
\begin{eqnarray}
(A \circ B)\circ C  &=& A\circ C + B\circ C - (A\cdot B)\circ C.
\label{prelemma1}
\end{eqnarray}
This equation will be used to prove lemma~1.

Now, it can be easily shown that the associativity of the circle-product arises from
that of the multiplication rule.\footnote{%
	Note further that if we had defined the circle-operator with the opposite sign-convention,
	$A \circ B = A\cdot B - A - B$,
	associativity would have failed.
%
\begin{eqnarray*}
(A \circ B)\circ C  &\stackrel{?}{=}& A\circ (B\circ C)\\
(A\cdot B- A - B)\circ C  &\stackrel{?}{=}& A\circ (B \cdot C -B - C  )\\
(A\cdot B- A - B)\cdot C - (A\cdot B- A - B) - C 
	&\stackrel{?}{=}& 
	A\cdot (B \cdot C -B - C  ) - A - (B \cdot C -B - C  )\\
A\cdot B\cdot C- A\cdot C - B\cdot C\mbox{\qquad\qquad}\\ - A\cdot B + A + B - C 
	&\stackrel{?}{=}& 
	A\cdot B \cdot C -A\cdot B - A\cdot C \\& &\mbox{\quad} - A - B \cdot C +B + C  \\
A - C
&\neq & 
	 - A + C
\end{eqnarray*}
}%
%
\begin{eqnarray*}
(A \circ B)\circ C  &\stackrel{?}{=}& A\circ (B\circ C)\\
(A + B - A\cdot B)\circ C  &\stackrel{?}{=}& A\circ (B + C - B \cdot C)\\
A\circ C + B\circ C - (A\cdot B)\circ C &\stackrel{?}{=}& 
	A\circ B + A\circ C - A\circ (B \cdot C)\\
(B+C-B\cdot C)\mbox{\qquad\qquad\qquad\qquad} \\
- (A\cdot B + C - (A\cdot B)\cdot C )
&\stackrel{?}{=}& 
(A+B -A\cdot B) 
\\
& &\mbox{\quad }
 - (A+ B \cdot C- A\cdot(B \cdot C)) \\
(A\cdot B)\cdot C
&\stackrel{\surd}{=}& 
	 A\cdot(B \cdot C).
\end{eqnarray*}

We now derive an algebraic identity which underlies the generalized interference term.\\

\noindent{\bf Lemma~3.} 
For mutually-disjoint subsets
$S_1, S_2, \cdots, S_n \in {\cal P(S)}$, the circle-product can be expressed as
\begin{eqnarray}
\lefteqn{
S_1 \circ S_2 \circ \ldots \circ S_n  =} \nonumber\\
&&\sum_{i=1}^n S_i 
-
\sum_{1\leq i< j}^n S_i \cdot S_j
+
\sum_{1\leq i<j<k}^n  S_i \cdot S_j \cdot S_k
\mp \ \ldots\ \nonumber\\
&& \qquad\qquad
-(-1)^{n-1}\sum_{i=1}^n S_1 \cdot  S_2 \cdot \ldots \cdot  \stackrel{omit}{S_i}\cdot \ldots \cdot S_n \nonumber\\
& & \qquad\qquad - (-1)^n  S_1\cdot S_2 \cdot \ldots \cdot S_n.
\label{lemma3}
\end{eqnarray}

\noindent{\bf Proof.}

Consider a pair of disjoint subsets $A,B\in {\cal P(S)}$. 
For $n=2$, the lemma is true by definition.
\begin{eqnarray*}
A\circ B  &=& A + B - A\cdot B.
\end{eqnarray*}
By forming 1 minus the left-hand side, 
observe that 
\begin{eqnarray*}
     1-A\circ B  
     &=& 1- (A + B - A\cdot B)\\
     &=& (1- A)\cdot (1 - B).
\end{eqnarray*}
In fact, 
for a set of $n$ mutually disjoint subsets  $A,B,C,\ldots,N\in {\cal P(S)}$,
we have
\begin{eqnarray}
     1-A\circ B\circ C \circ\ldots\circ N 
     &=& 1-A\circ (B\circ C \circ\ldots\circ N) \nonumber\\
     &=& (1-A)\cdot ( 1-(B\circ C \circ\ldots\circ N) ) \nonumber\\
     &=& (1-A)\cdot \left[ (1-B)\cdot (1-(C \circ\ldots\circ N)) \right] \nonumber\\
     &=& (1-A)\cdot \left[(1- B)\cdot(1- C)\cdot\dots \cdot(1 - N) \right] \nonumber\\
     &=& (1- A)\cdot (1 - B)\cdot(1-C)\cdot\ldots\cdot(1-N)
\label{lemma3proof}     
\end{eqnarray}
By expanding out the right-hand side of Eq.~(\ref{lemma3proof}), we obtain
1 minus the right-hand side of Eq.~(\ref{lemma3}).
\hfill\qed

\subsection{Duality}\label{prelemma2}
\noindent
We note the following duality between the multiplication rule and the circle-product.
With simple algebra, the definition of the circle-product can be reversed to read
$$ A \cdot B := A +B -A\circ B.$$  Formally, it appears that one can swap the roles
of ``~$\cdot$~'' and ``$\circ$'' in a valid equation and obtain another valid equation.
Let us make this more precise.

For any $A\in Z{\cal P(S)}$, define its dual to be $A':=\EMPTYSET - A$. 
Note that $\EMPTYSET'=0$ and $0'=\EMPTYSET$. Clearly,  we have $(A')'=A$.

Consider $C=A\circ B$.
\begin{eqnarray*}
C
&=&A\circ B\\
&=&(\EMPTYSET - A')\circ (\EMPTYSET - B')\\
&=&(\EMPTYSET - A')+ (\EMPTYSET - B') - (\EMPTYSET - A')\cdot (\EMPTYSET - B')\\
&=&2 \EMPTYSET - A'- B' - \EMPTYSET^2 +\EMPTYSET\cdot B' + A'\cdot\EMPTYSET - A'\cdot B'\\
&=&2 \EMPTYSET - A'- B' - \EMPTYSET + B'+ A' - A'\cdot B'\\
\EMPTYSET-C'&=& \EMPTYSET - A'\cdot B'.
\end{eqnarray*}
So, we find that
\begin{eqnarray*}
(A\circ B)'&=&A'\cdot B'.
\end{eqnarray*}
A similar calculation verifies that
\begin{eqnarray*}
(A\cdot B)'&=&A'\circ B'.
\end{eqnarray*}

Now consider $C=A+B$, then
\begin{eqnarray*}
C
&=&A + B\\
&=&(\EMPTYSET - A') + (\EMPTYSET - B')\\
\EMPTYSET - C'&=&2\EMPTYSET - A' - B'.
\end{eqnarray*}
So,
\begin{eqnarray*}
(A + B)'&=&A' + B' - \EMPTYSET.
\end{eqnarray*}
Thus, for general sums, the duality-operation does not distribute over addition.
However, the duality-operation does distribute over affine sums.
\begin{eqnarray*}
\left(\sum n_i S_i\right)' &\stackrel{?}{=}& \sum n_i\left(  {S_i}' \right)\\
\EMPTYSET-\sum n_i S_i &\stackrel{?}{=}& \sum n_i\left(  \EMPTYSET-S_i \right)\\
	&\stackrel{?}{=}& \sum n_i \EMPTYSET-\sum n_i S_i \\
\EMPTYSET &\stackrel{?}{=}& \left(\sum n_i\right) \EMPTYSET\\
1 &\stackrel{\surd}{=}&  \left(\sum  n_i\right).
\end{eqnarray*}

\break

	Without proof, we state the following duality theorem.\\

\noindent{\bf Duality Theorem}\\
Let $S_1$ and $S_2$ be expressions formed from $0$, $\EMPTYSET$ and 
indeterminates $A$, $B$, $C$, etc., using 
``~$\cdot$~'', ``$\circ$'' and affine linear combination.
Let ${S_1}'$ and ${S_2}'$ be their dual expressions 
obtained by swapping all occurrences of ``$\circ$'' with ``~$\cdot$~'' and
of $\EMPTYSET$ with $0$.   
If $S_1=S_2$ is an identity in $Z{\cal P(S)}$,
then ${S_1}'={S_2}'$ is also an identity in $Z{\cal P(S)}$.\\

These results will be used to prove lemma~2.

\subsection{The Extended Generalized Measure}
\noindent
Consider a linear map, which we will call the ``Extended Generalized Measure,''
$\mu: Z{\cal P(S)} \rightarrow {\bf R}$, where ${\bf R}$ denotes
the real numbers.

Consider a pair of disjoint subsets $A,B \in {\cal P(S)}$.
Applying this map to the circle-product of $1A,1B \in Z{\cal P(S)}$,
we have:
\begin{eqnarray}
A\circ B &=& A + B - A\cdot B\nonumber\\
         &=& A + B - (A\sqcup B)\nonumber\\
\mu(A\circ B) &=& \mu(A) + \mu(B) - \mu(A\sqcup B)\label{MU}.
\end{eqnarray}

In order to make the connection with the Generalized Measure $|\cdot|$ as 
defined by Sorkin,\cite{QMT} let us impose the following
conditions on $\mu$.  We require that $\mu(\EMPTYSET)=0$
and that $\mu(1X)\geq 0$ for all $X\in {\cal P(S)}$.
In other words, we require that $\mu$ be a
Generalized Measure extended to $Z{\cal P(S)}$ by linearity.

Now, let us notice the following. 
For mutually-disjoint
$S_1, S_2, \cdots, S_n \in {\cal P(S)}$, we have
\begin{eqnarray*}
I_n(S_1, S_2 , \cdots, S_i, \cdots, S_n)
&=&
\mu( \bigcirc_{i=1}^{n} S_i ).
\end{eqnarray*}
So, Eq.~(\ref{MU}) can be written
\begin{eqnarray*}
I_2(A,B) &=& I_1(A) + I_1(B) - I_1(A\sqcup B),
\end{eqnarray*}
which agrees (up to an overall sign) with the definition given by 
Sorkin.\cite{QMT}
Similarly, the corresponding higher-order expressions clearly agree for all $n$.

We are now prepared to give proofs of lemma~1 and lemma~2.

\pagebreak

\section{Proofs of the lemmas\label{PROOF}}
\noindent{\bf Lemma~1.~\cite{QMT}}
\begin{eqnarray*}
\lefteqn{I_n(A,B,C,\cdots,N)=}&&\\
&&
I_{n-1}(A, C, \cdots, N)+
I_{n-1}(B, C, \cdots, N)
-I_{n-1}(A\sqcup B, C, \cdots, N).
\end{eqnarray*}

\noindent{\bf Proof.}
First, consider a collection of three mutually disjoint subsets
$A,B,X\in {\cal P(S)}$.
Forming the triple circle-product in $Z{\cal P(S)}$, 
$$(A\circ B) \circ X = (A + B - A\cdot B) \circ X,$$
and using the associativity and affine-distributivity of the circle-product (see Eq.~(\ref{prelemma1})), we have
\begin{eqnarray*}
(A\circ B) \circ X &=& A\circ X + B\circ X - (A\cdot B) \circ X\\
A\circ B \circ X &=& A\circ X + B\circ X - (A\sqcup B) \circ X,
\end{eqnarray*}
where we used the definition of multiplication on the right-hand side.
Applying the linear map $\mu$ and making the identifications defined
in the last section, we find
\begin{eqnarray*}
\mu(A\circ B \circ X) 
&=& \mu(A\circ X) + \mu(B\circ X) - \mu((A\sqcup B) \circ X)\\
I_3(A, B, X) 
&=& I_2(A,X)+ I_2(B,X) -I_2(A\sqcup B, X).
\end{eqnarray*}

By taking $1X \in Z{\cal P(S)}$ to be of the form
\begin{eqnarray*}
X
&=& \bigcirc_{i=1}^{n-2} X_i \ =\ C\circ D\circ \ldots \circ N,
\end{eqnarray*}
where $A,B,C,D,\cdots,N \in {\cal P(S)}$ are mutually disjoint,
we complete the proof for all $n>1$:
\begin{eqnarray}
A\circ B \circ X &=& A\circ X + B\circ X - (A\cdot B) \circ X\\
&&\nonumber\\
\mu(A\circ B \circ X) 
&=& \mu\left(A\circ X\right) + \mu\left(B\circ X\right) - \mu\left((A\sqcup B) \circ X\right)
	\nonumber\\
\mu\left(A\circ B \circ \left(\bigcirc_{i=1}^{n-2} X_i \right) \right) 
&=& \mu\left(A\circ \left(\bigcirc_{i=1}^{n-2} X_i \right)\right) 
+ \mu\left(B\circ \left(\bigcirc_{i=1}^{n-2} X_i \right)\right) \nonumber \\
&&\qquad  - \mu\left((A\sqcup B) \circ \left(\bigcirc_{i=1}^{n-2} X_i \right)\right) \nonumber\\
I_n(A,B,C, \cdots, N)
&\stackrel{\surd}{=}& I_{n-1}(A,C, \cdots, N)
+ I_{n-1}(B,C, \cdots, N) \nonumber\\
&&\qquad  - I_{n-1}(A\sqcup B,C, \cdots, N).\nonumber
\end{eqnarray}
\hfill\qed

\pagebreak

\noindent{\bf Lemma~2.}
\begin{eqnarray*}
\lefteqn{
\left|S_1\sqcup S_2  \sqcup \cdots \sqcup S_n \right|=}\\
&&\sum_i I_1(S_i)
-
\sum_{\mbox{distinct }i,j} I_2(S_i,S_j)
+
\sum_{\mbox{distinct }i,j,k} I_3(S_i,S_j,S_k)
\mp \ \ldots\ \\
&& \qquad\qquad
-(-1)^{n-1}\sum_i I_{n-1}(S_1, S_2,\cdots, \stackrel{omit}{S_i}, \cdots, S_n)\\
& & \qquad\qquad - (-1)^n  I_n(S_1, S_2, \cdots, S_n).
\end{eqnarray*}

\noindent{\bf Proof.}
	Consider lemma 3.
For mutually-disjoint subsets
$S_1, S_2, \cdots, S_n \in {\cal P(S)}$, the circle-product can be expressed as
\begin{eqnarray*}
\lefteqn{
S_1 \circ S_2 \circ \ldots \circ S_n  =}\\
&&\sum_{i=1}^n S_i
-
\sum_{1\leq i< j}^n S_i \cdot S_j
+
\sum_{1\leq i<j<k}^n  S_i \cdot S_j \cdot S_k
\mp \ \ldots\ \\
&& \qquad\qquad
-(-1)^{n-1}\sum_{i=1}^n S_1 \cdot  S_2 \cdot \ldots \cdot  \stackrel{omit}{S_i}\cdot \ldots \cdot S_n\\
& & \qquad\qquad - (-1)^n  S_1\cdot S_2 \cdot \ldots \cdot S_n.
\end{eqnarray*}

	So, applying the duality-operation%
\footnote{%
	Note that each term of the form $S_1\cdot S_2\cdot \ldots\cdot S_m$ has coefficient equal to $1$. 
	The sum of the coefficients of these terms 
	is $-\sum_{i=1}^n (-1)^i \! \left( \hspace*{-1.5ex}\begin{array}{c} n\\i \end{array}\hspace*{-1.5ex}\right)$.
	Using $(1+x)^n=\sum_{i=0}^n (-1)^i \!\left(\hspace*{-1.5ex} \begin{array}{c} n\\i \end{array} \hspace*{-1.5ex}\right)x^n$
	with $n>0$ and $x=-1$, we find that the sum of the coefficients is equal to $1$.
	}~~ 
of Section~4 
to lemma~3, we obtain

\begin{eqnarray*}
\lefteqn{
S_1 \cdot S_2 \cdot \ldots \cdot S_n  =}\\
&&\sum_{i=1}^n S_i
-
\sum_{1\leq i< j}^n S_i \circ S_j
+
\sum_{1\leq i<j<k}^n  S_i \circ S_j \circ S_k
\mp \ \ldots\ \\
&& \qquad\qquad
-(-1)^{n-1}\sum_{i=1}^n S_1 \circ  S_2 \circ \ldots \circ  \stackrel{omit}{S_i}\circ \ldots \circ S_n\\
& & \qquad\qquad - (-1)^n  S_1\circ S_2 \circ \ldots \circ S_n.
\end{eqnarray*}

Writing ``$1(S_1\sqcup S_2  \sqcup \cdots \sqcup S_n)$'' for $S_1 \cdot S_2 \cdot \ldots \cdot S_n$ 
and then applying the linear map $\mu$, we obtain the statement of lemma 2.
\hfill\qed

\section{Open Questions}

We have found an interesting algebraic structure underlying the
Quantum Measure and its generalizations.  What is its physical
interpretation for both classical and quantum physics?

By themselves, the sum rules for the ``$I_3=0$'' case do not 
uniquely yield the standard quantum theory.
What additional axioms are needed to select the standard quantum theory
from all possible ``$I_3=0$'' theories?

In this paper, we were mainly concerned with the special case of
mutually disjoint subsets, which is sufficient to prove these identities 
involving the Quantum Measure.
Sorkin\cite{QMT} showed how one could extend the definition of $I_2$
to general (non-disjoint) arguments, which he used to show
that every Quantum Measure [which satisfies the ''$I_3=0$'' sum rule]
can arise in this way. Can a similar extension to general arguments
be carried out for the higher-order interference functions $I_n$?

In particular, we showed that a probability function $P_r$ that assigns
to any disjoint union of subsets the $r^{th}$-power of its amplitude
satisfies the ``$I_n=0$'' sum rule if $r<n$.  Is the converse true?
Or are there other functional relationships between the probabilities 
and the amplitudes that satisfy these sum rules?

Sorkin\cite{QMT} has proposed a ``null'' three-slit experiment to test the
validity of standard quantum mechanics.  A non-null result would indicate
that a more general dynamics was at work.  In light of this possibility,
can an ``$I_4=0$'' generalization of quantum mechanics be formulated?

\nonumsection{Acknowledgments}
\noindent
This paper is based on a seminar delivered at 
UNAM-ICN, Mexico City, May 1996.
I would like to thank Rafael Sorkin and David Rideout 
for many useful comments.
I especially thank Rafael Sorkin for
suggesting a simpler proof of Lemma~3.

This research was partly supported by NSF grant PHY-9600620 
and by a grant from the Office of Research and Computing of 
Syracuse University.

\appendix
\noindent
In this section, we prove the associative and distributive
properties of the multiplication rule $\cdot: Z{\cal P(S)} \times Z{\cal P(S)} \rightarrow Z{\cal P(S)}$.
To simplify the notation used in this proof, we will write
\begin{eqnarray}
\left( \sum_i a_i S_i\right)\cdot \left( \sum_j b_j S_j\right)
&:=&\left( \sum_k [ab]_k S_k\right),
\end{eqnarray}
where $a_i, b_i \in Z$ and $S_i \in {\cal P(S)}$, with
\begin{eqnarray*}
[ab]_k=
\displaystyle \sum_{  \stackrel{i,j} { S_i\sqcup S_j = S_k } }
					  \hspace{-2ex}a_i b_j,
\end{eqnarray*}
with the understanding that we take $[ab]_k=0$ if the summation is empty.
We will also use the summation convention by writing $a_i S^i$ 
for $\sum_i a_i S_i$.

First, we show distributivity over addition.
\begin{eqnarray*}
(a_i S^i) \cdot \left((b_j +  c_j) S^j\right)
&\stackrel{?}{=}&
  \left(a_i S^i \cdot b_jS^j \right) 
+   \left(a_iS^i \cdot  c_jS^j \right)\\
{[a(b+ c)]}_m S^m  
&\stackrel{?}{=}&
[ab]_m S^m
+ [ a c]_m S^m\\
{[a(b+ c)]}_m
&\stackrel{?}{=}&
[ab]_m
+ [a  c]_m\\
\sum_{  \stackrel{i,j} { S_i\sqcup S_j = S_m } }
					  \hspace{-2ex}a_i (b_j+ c_j)
&\stackrel{?}{=}&
\sum_{  \stackrel{i,j} { S_i\sqcup S_j = S_m } }
					  \hspace{-2ex}a_i b_j
+
\sum_{  \stackrel{i,j} { S_i\sqcup S_j = S_m } }
					  \hspace{-2ex}a_i  c_j\\
\sum_{  \stackrel{i,j} { S_i\sqcup S_j = S_m } }
					  \hspace{-2ex}a_i b_j
+
\sum_{  \stackrel{i,j} { S_i\sqcup S_j = S_m } }
					  \hspace{-2ex}a_i  c_j
&\stackrel{\surd}{=}&
\sum_{  \stackrel{i,j} { S_i\sqcup S_j = S_m } }
					  \hspace{-2ex}a_i b_j
+
\sum_{  \stackrel{i,j} { S_i\sqcup S_j = S_m } }
					  \hspace{-2ex}a_i  c_j ,
\end{eqnarray*}
where we have used the distributivity of ordinary multiplication in $Z$.

Finally, we show associativity.
\begin{eqnarray*}
\left( (a_i S^i) \cdot (b_j S^j)\right) \cdot (c_k S^k)
&\stackrel{?}{=}&
 (a_i S^i) \cdot \left( (b_j S^j) \cdot (c_k S^k) \right)\\
\left( [ab]_m S^m \right) \cdot (c_k S^k)
&\stackrel{?}{=}&
 (a_i S^i) \cdot \left( [bc]_n S^n \right)\\
{[[ab]c]}_p S^p
&\stackrel{?}{=}&
 {[a[bc]]}_q S^q\\
{[[ab]c]}_p
&\stackrel{?}{=}&
 {[a[bc]]}_p\\
\sum_{  \stackrel{m,k} { S_m\sqcup S_k = S_p } }
\left(\sum_{  \stackrel{i,j} { S_i\sqcup S_j = S_m } }
					  \hspace{-2ex}a_i b_j \right) c_k
&\stackrel{?}{=}&
\sum_{  \stackrel{i,n} { S_i\sqcup S_n = S_p } } a_i
\left(\sum_{  \stackrel{j,k} { S_j\sqcup S_k = S_n } }
					  \hspace{-2ex}b_j c_k \right)\\
\sum_{  \stackrel{m,k} { S_m\sqcup S_k = S_p } }\
\sum_{  \stackrel{i,j} { S_i\sqcup S_j = S_m } }
					  \hspace{-2ex}a_i b_j  c_k
&\stackrel{?}{=}&
\sum_{  \stackrel{i,n} { S_i\sqcup S_n = S_p } }\
\sum_{  \stackrel{j,k} { S_j\sqcup S_k = S_n } }
					  \hspace{-2ex} a_i b_j c_k\\
\sum_{  \stackrel{i,j,k} { (S_i\sqcup S_j)\sqcup S_k = S_p } }\
					  \hspace{-2ex}a_i b_j  c_k
&\stackrel{\surd}{=}&
\sum_{  \stackrel{i,j,k} { S_i\sqcup (S_j\sqcup S_k) = S_p } }\
					  \hspace{-2ex} a_i b_j c_k .
\end{eqnarray*}
where we have used the associativity of the disjoint-union $\sqcup$
in ${\cal P(S)}$.

This completes the proof that $(Z{\cal P(S)},+,\cdot)$ is a ring.

\nonumsection{References}

\end{document}